\documentclass[11pt, oneside]{article} 
\usepackage[utf8]{inputenc}
\usepackage[utf8]{inputenc}
\usepackage{geometry}                		
\usepackage{graphicx}				
\usepackage{amssymb}
\usepackage{amsmath}
\usepackage{subfig}
\usepackage{caption}
\usepackage{comment}
\usepackage{amsthm,verbatim,amsfonts,amscd,float,physics}
\usepackage{appendix}

\usepackage[utf8]{inputenc}
\usepackage{xcolor}
\numberwithin{equation}{section}
\usepackage{titling, lipsum}
\title{Superconformal Quantum Mechanics\\ on a Quantum Computer}
\author{Yuan Feng$^{(1)}$, Michael McGuigan, Tristen White$^{(2)}$\\
(1) University of California, Berkeley\\ 
(2) Florida State University\\
email contact: michael.d.mcguigan@gmail.com\\
}
\date{}

\begin{document}
\begin{titlingpage}
\maketitle

\begin{abstract}
We investigate  superconformal quantum mechanics (SCQM) on a quantum computer. We study the ground state of the mass deformed SCQM on a quantum computed using the Variational Quantum Eigensolver (VQE) using a one boson and one boson - one fermion Hilbert space with and without noise and compare the accuracy of the results. We study the Feynman path integral for SCQM using the Evolution of Hamiltonian (EOH) algorithm on the quantum computer using the Trotter-Suzuki approximation and compare with the exact result.  We consider an N boson and N fermion version of SCQM given be the Supersymmetric Calogero-Moser-Sutherland (SCMS) model. We compare the ground state of SCMS theory obtained using the VQE computation with the exact solution. Finally we discuss the implications of the the numerical simulation of SCQM on a quantum computer for the simulation of quantum gravity in light of the Anti-de Sitter/Superconformal field theory (AdS/CFT) correspondence.   
\end{abstract}
\end{titlingpage}
\newpage

\section{Introduction}

Superconformal Quantum  Mechanics (SCQM) is part of a family of supersymmmetric Hamiltonians such as Hamiltonians  with quartic potentials  in 3+1 dimensions and sixth order potentials 2+1 dimensions. Superconformal Quantum Mechanics  has a inverse power of two potential in 0+1 dimensions \cite{deAlfaro:1976vlx}\cite{Akulov:1983hjq}\cite{Fubini:1984hf}\cite{Freedman:1990gd}\cite{Desrosiers:2003me}. We will consider  the  mass deformed version of the potential that has an addition quadratic term in the potential. Superconformal Quantum Mechanics has interesting applications including the study of Baryon Spectrum in the light front formalism \cite{deTeramond:2014asa}\cite{Dosch:2015nwa}
\cite{Dosch:2020hqm}, black holes physics \cite{Gaiotto:2004ij}\cite{Michelson:1999dx}\cite{Lechtenfeld:2015wka},
relations to $AdS_2$ and holography \cite{Astorino:2002bj}
\cite{Verlinde:2004gt} and also has application to quantum cosmology \cite{Pioline:2002qz}. 

In this paper we study Superconformal Quantum mechanics using quantum computing. Quantum computing is a potentially disruptive form of computing that may excel at the simulation of quantum systems. As superconformal quantum mechanics is exactly soluble it can serve as a benchmark for scalability and accuracy of quantum computation in regimes beyond which can be verified using classical computers. In this paper we shall study quantum computations up to 9 qubits but the methods can in principle scaled much further and one should keep in mind that quantum computers already exist with capacities exceeding 100 qubits. We will use different quantum algorithms in our simulation of superconformal quantum mechanics such as the Variational Quantum Eigensover (VQE) and the Evolution of Hamiltonian (EOH) quantum algorithm. We study different bases to represent the Hamiltonian on the quantum computer and compare the accuracy with that of the classical computer. Finally we extend our analysis to the multi-particle supersymmetric  Calogero-Moser-Sutherland model which may have applications to the study of Yang Mills Theories \cite{Fedoruk:2010jwv} \cite{Gorsky:1993pe}, holographic quantum gravity \cite{McGreevy:2003dn} and many body physics \cite{Calogero:1975ii}.

\section{Superconformal Quantum Mechanics}

For superconformal quantum mechanics with mass deformation the superpotential is taken to be:
\begin{equation}W(x) =m \omega x - \frac{g}{x}\end{equation}
Then the minus and plus partner potentials are defined as \cite{Witten:1981nf} \cite{Cooper:1994eh}
\cite{Gangopadhyaya:2011wka}:
\begin{equation}\begin{array}{l}
{V_ - }(x) = {W^2}(x) - W'(x)\\
{V_ + }(x) = {W^2}(x) + W'(x)
\end{array}\end{equation}
and are given by:
\[{V_ - }(x) =  m^2 \omega^2 {x^2} + (g^2-g) {x^{ - 2}} - 2 m \omega g  - m\omega \]
\begin{equation}{V_ + }(x) =m^2 \omega^2{x^2} + (g^2+g){x^{ - 2}} - 2 m \omega g  + m \omega
\end{equation}
The partner Hamiltonians are
\[{H_ - } = {p^2} + {V_ - }(x)\]
\begin{equation}{H_ + } = {p^2} + {V_ + }(x)\end{equation}
Rescaling the minus partner Hamiltonian by $\frac{1}{2m}$ we define the Hamiltonian $H$ as:
\begin{equation} H = \frac{1}{2m} H_{-} = \frac{1}{2m} ( {p^2} + {V_ - }(x))\end{equation}
For $m =\frac{1}{2}$, $\omega =1$, $g = \sqrt 2$ the ground state of the minus partner Hamiltonian is:
\begin{equation}\psi _0^ -  = {e^{ - \int_{{x_0}}^x {W(x)dx} }} = \frac{1}{2}{\left( x \right)^{\sqrt 2 }}{e^{ - \frac{1}{4}{x^2}}}\end{equation}
and the energy spectrum of the partner Hamiltonians are:
\[\varepsilon _n^ -  = 0,2,4,6,8, \ldots \]
\begin{equation}\varepsilon _n^ +  = 2,4,6,8, \ldots \end{equation}
\begin{figure}
\centering
  \includegraphics[width = .5 \linewidth]{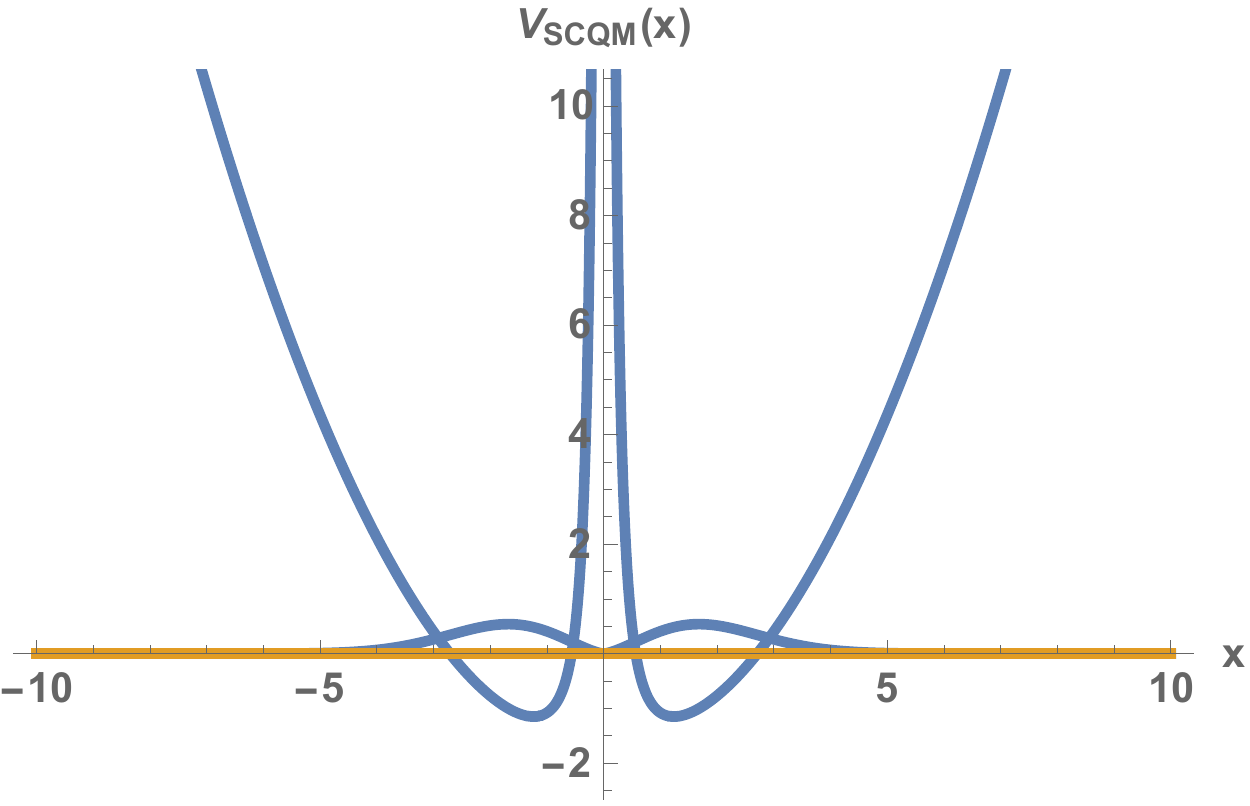}
  \caption{Supersymmetric Conformal Quantum Mechanics (SCQM) potential and ground state wave function for $g=\sqrt{2}$. }
  \label{fig:Radion Potential}
\end{figure}
The potential and ground state wave function is plotted in figure 1. One can form $A$ and $A^{\dagger}$ operators through
\[A = \frac{1}{{\sqrt {2 m} }}(ip + m \omega x - g {x^{ - 1}})\]
\begin{equation}{A^\dag } = \frac{1}{{\sqrt {2 m} }}( - ip + m \omega x - g {x^{ - 1}})\end{equation}
Then the Hamiltonian is simply:
\begin{equation}{H} = {A^\dag }A\end{equation}
This form will be used in our quantum computations and turns out to yield highly accurate results.

\section{Quantum computing basis representations}

Quantum computing for supersymmetric models has recently attracted  interest because of their many mathematically and physically interesting features \cite{Rinaldi:2021jbg}\cite{Apanavicius:2021yin}\cite{Miceli}\cite{Kane}\cite{Culver:2021rxo}. In quantum computing one represents the quantum operators as finite matrices in a given basis. In this paper we study the oscillator basis, position basis and finite difference basis described below:

Once we have operators represented as matrices one performs an expansion of these matrices in terms of tensor products of the three Pauli matrices and the identity matrix called Pauli terms in order to represent them on a quantum computer.The quantum algorithms we will use in this paper are the Variational Quantum Eigensolver (VQE) to estimate  the ground state energy and wave function and the Evolution of Hamiltonian (EOH) algorithm to estimate the Kernel or path integral. 

\subsection*{Gaussian or Simple Harmonic Oscillator basis}

This is a very useful basis based on the matrix treatment of the simple harmonic oscillator which is sparse in representing the position and momentum operator. For the position operator we have:
\begin{equation} 
 X_{osc} = \frac{1}{\sqrt{2}}\begin{bmatrix}
 
   0 & {\sqrt 1 } & 0 &  \cdots  & 0  \\ 
   {\sqrt 1 } & 0 & {\sqrt 2 } &  \cdots  & 0  \\ 
   0 & {\sqrt 2 } &  \ddots  &  \ddots  & 0  \\ 
   0 & 0 &  \ddots  & 0 & {\sqrt {N-1} }  \\ 
   0 & 0 &  \cdots  & {\sqrt {N-1} } & 0  \\ 
\end{bmatrix}
  \end{equation}
while for the momentum operator we have:
\begin{equation}
 P_{osc} = \frac{i}{\sqrt{2}}\begin{bmatrix}
 
   0 & -{\sqrt 1 } & 0 &  \cdots  & 0  \\ 
   {\sqrt 1 } & 0 & -{\sqrt 2 } &  \cdots  & 0  \\ 
   0 & {\sqrt 2 } &  \ddots  &  \ddots  & 0  \\ 
   0 & 0 &  \ddots  & 0 & -{\sqrt {N-1} }  \\ 
   0 & 0 &  \cdots  & {\sqrt {N-1} } & 0  \\ 
\end{bmatrix}
  \end{equation}
The  SCQM  Hamiltonian $H_{-}$ is then 
\begin{equation}
 H_{-}=P_{osc}^2 + m^2\omega^2 X_{osc}^2 +(g^2-g ) X_{osc}^{-2} - (m\omega + g 2 m \omega)  I   
\end{equation}
where  $I$ is the $N \times N$ identity matrix.

\subsection*{Position basis}

In the position basis the position matrix is diagonal but the momentum matrix is dense and constructed from the position operator using a Sylvester matrix $F$. In the position basis the position matrix is:
\begin{equation}
{\left( {{X_{pos}}} \right)_{j,k}} = \sqrt {\frac{{2\pi }}{{4N}}} (2j - (N + 1)){\delta _{j,k}}
\end{equation}
and the momentum matrix is:
\begin{equation}{P_{pos}} = {F^\dag }{X_{pos}}F\end{equation}
where 
\begin{equation}{F_{j,k}} = \frac{1}{{\sqrt N }}{e^{\frac{{2\pi i}}{{4N}}(2j - (N + 1))(2k - (N + 1))}}\end{equation}
The SCQM Hamiltonian is  then
\begin{equation}
 H_{-}=P_{pos}^2 + m^2\omega^2 X_{pos}^2 +(g^2-g ) X_{pos}^{-2} - (m\omega + g 2 m \omega)  I  
\end{equation}
and in this case the matrix potential is very simple as it is a function of a diagonal matrix.

\subsection*{Finite difference basis}

This is the type of basis that comes up when realized differential equations in terms of finite difference equations. In this case the position operator is again diagonal but the momentum operator although not diagonal is still sparse. In the finite difference basis the position matrix is:
$${\left( {{X_{fd}}} \right)_{j,k}} = \sqrt {\frac{1}{{2N}}} (2j - (N + 1)){\delta _{j,k}}$$
and the momentum-squared matrix is:
\begin{equation} 
 P_{fd}^2 = \frac{N}{2}\begin{bmatrix}
 
   2 & - 1  & 0 &  \cdots  & 0  \\ 
   -1 & 2 & -1 &  \cdots  & 0  \\ 
   0 & -1 &  \ddots  &  \ddots  & 0  \\ 
   0 & 0 &  \ddots  & 2 & -1  \\ 
   0 & 0 &  \cdots  & -1 & 2  \\ 
\end{bmatrix}
  \end{equation}
The SCQM Hamiltonian is then:
\begin{equation}
  H_{-}=P_{fd}^2 + m^2\omega^2 X_{fd}^2 +(g^2-g) X_{fd}^{-2} - (m\omega + g 2 m \omega)  I  
\end{equation}

\subsection*{Annihilation and Creation operator basis}

The $A$ operator basis is similar to the oscillator basis. One forms
\[A = \frac{1}{{\sqrt {2 m} }}(i P_{osc} + m \omega X_{osc} - g {X_{osc}^{ - 1}})\]
\begin{equation}{A^\dag } = \frac{1}{{\sqrt {2 m} }}( - i P_{osc} + m \omega X_{osc} - g {X_{osc}^{ - 1}})\end{equation}
Then the Hamiltonian is simply:
\begin{equation}{H} = {A^\dag }A\end{equation}
Whatever basis one uses one needs to map the Hamiltonian to a an expression in terms of a sum of tensor products of Pauli spin matrices $\sigma_1, \sigma_2,\sigma_3$ plus the identity matrix which are the Pauli terms. As there are four such matrices the maximum number of terms in this expansion is $4^n$  where $n$ is the number of qubits. In most of our simulations the number of qubits was fixed at 4 so that the maximum number of Pauli terms was 256.

\section{Quantum computing results - One boson formulation}

The one boson formulation is the traditional formulation used in supersymmetric quantum mechanics. In this section we use the aop basis and write the Hamiltonian as:
\begin{equation}H = \frac{1}{2 m}\left( { - ip +  m \omega x - g {x^{ - 1}}} \right)\left( {ip +m \omega  x - g {x^{ - 1}}} \right)\end{equation}
for $g= \sqrt 2$.
The size of the matrix we used was $16\times 16$ so the Hamiltonian in the one boson formulation used 4 qubits. We use the Variational Quantum Eigensolver (VQE) quantum algorithm which is part of IBM Qiskit, a toolkit developed by IBM to run quantum algorithms on quantum computers and simulators.. The VQE quantum algorithm works by representing a variational quantum state in terms of rotation gates parametrized by angles. Then one uses an optimizer to vary the angles from one iteration to the next to try to find the lowest energy which is determined by the algorithm to lie above the exact ground state energy. We used two types of simulations, one using the state-vector simulator and no noise, and another that used the QASM simulator with realistic noise. The results are shown in tables 1 and 2 with convergence graphs in figures 2 and 3. For the no noise simulation we found highly accurate results using the Limited-memory BFGS Bound (L-BFGS-B)  optimizer. For the noisy simulation not surprisingly the simulation was less accurate but still produced a ground state energy estimate close to zero. The accuracy of the no noise simulation was somewhat surprising given the relatively large value of the coupling. This means the parametrization of the ground state wave function in terms of parametrized rotation gates had a strong overlap with the exact ground state wave function. Previous studies of supersymmetric quantum mechanics showed a strong overlap with this type of parametrization only for weak coupling.
\begin{figure}
\centering
  \includegraphics[width = .5 \linewidth]{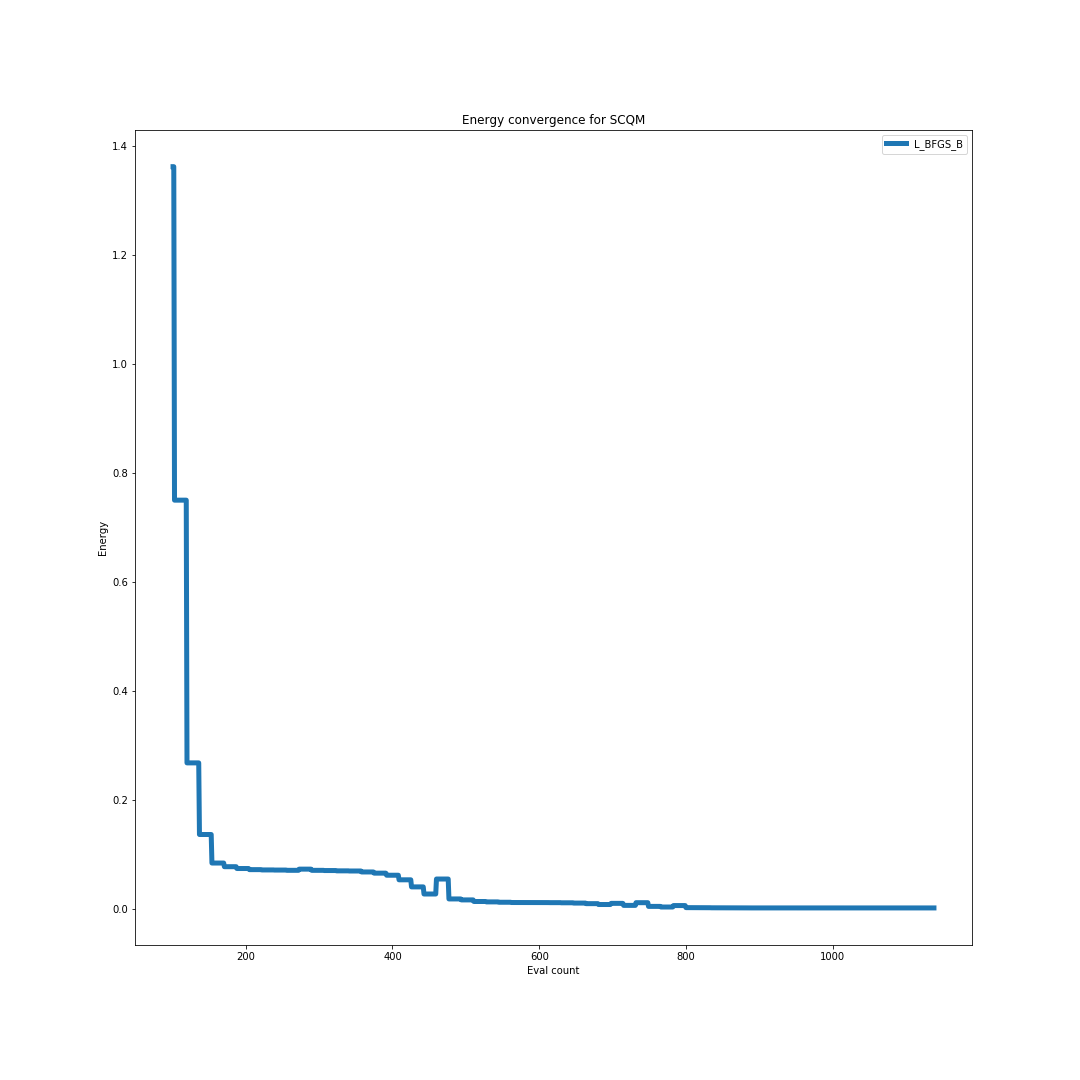}
  \caption{Convergence graph for the VQE computation of the ground state energy of SCQM using the one boson  formulation. The VQE result was obtained using the State-vector simulator with no noise and the Limited-memory BFGS Bound (L-BFGS-B)  optimizer.}
  \label{fig:Radion Potential}
\end{figure}
\begin{figure}
\centering
  \includegraphics[width = .5 \linewidth]{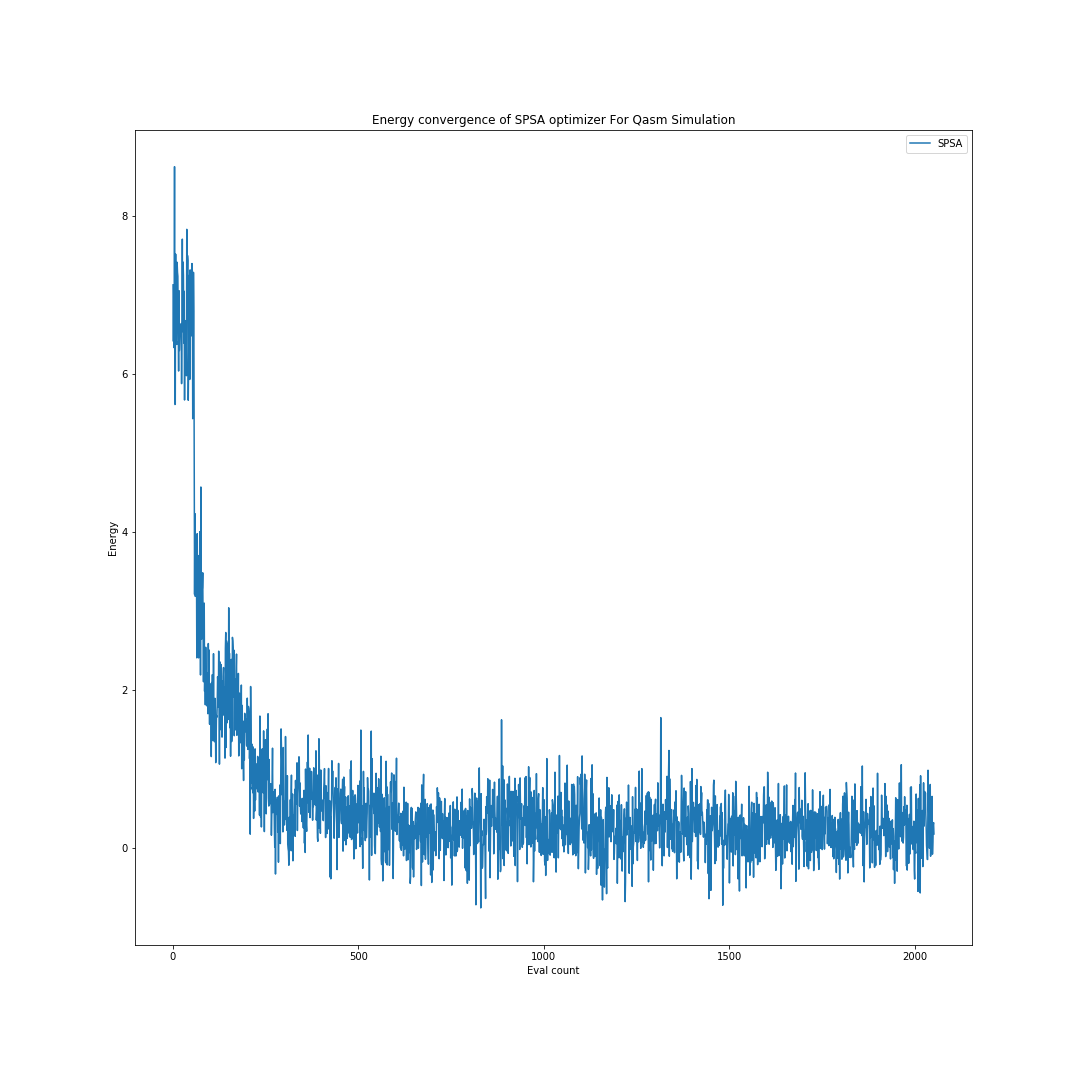}
  \caption{Convergence graph for the VQE computation of the ground state energy of SCQM using the one boson formulation. The VQE result was obtained using the QASM simulator including realistic noise and the Simultaneous Perturbation Stochastic Approximation (SPSA) optimizer.}
  \label{fig:Radion Potential}
\end{figure}
\begin{table}[h]
\centering
\begin{tabular}{|l|l|}
\hline
SCQM one boson formulation       & Ground State Energy \\ \hline
Exact   &    0.0        \\ \hline
Exact Discrete   &   0.001904019686       \\ \hline
VQE-State-vector (L-BFGS-B optimizer)   &    0.001904019785       \\ \hline
\end{tabular}
\caption{\label{tab:table-name} Exact, Exact discrete, and VQE computation of the ground state energy of SCQM using the one boson one fermion formulation. The VQE result was obtained using the State-vector simulator with no noise and the Limited-memory BFGS Bound (L-BFGS-B)  optimizer.}
\end{table}
\begin{table}[h]
\centering
\begin{tabular}{|l|l|}
\hline
SCQM one boson formulation       & Ground State Energy \\ \hline
Exact   &    0.0        \\ \hline
Exact Discrete   &   0.001904019686       \\ \hline
VQE-QASM-Simulator (SPSA optimizer)   &   0.178733653271       \\ \hline
\end{tabular}
\caption{\label{tab:table-name} Exact, Exact discrete, and VQE computation of the ground state energy of SCQM using the one boson one fermion formulation. The VQE result was obtained using the QASM simulator including realistic noise and the Simultaneous Perturbation Stochastic Approximation (SPSA) optimizer.}
\end{table}

\section{Quantum computing results - One boson and one fermion formulation}

The one boson one fermion formulation of supersymmetric conformal quantum mechanics is more in line with higher dimensional formulations of supersymmetric quantum field theory. In this case we the oscillator basis and use tensor products to define the boson-fermion Hilbert space from:
\[X = {X_{osc}} \otimes \left( {\begin{array}{*{20}{c}}
1&0\\
0&1
\end{array}} \right)\]
\[P = {P_{osc}} \otimes \left( {\begin{array}{*{20}{c}}
1&0\\
0&1
\end{array}} \right)\]
\begin{equation}C = {I_{16}} \otimes \left( {\begin{array}{*{20}{c}}
0&1\\
0&0
\end{array}} \right)\end{equation}
The supercharge is then defined by:
\begin{equation}{Q_{susy}} = C\left( {iP + X - \sqrt 2 {X^{ - 1}}} \right)\end{equation}
and the Hamiltonian is then:
\begin{equation}H = \frac{1}{2}\left( {{Q_{susy}}Q_{susy}^\dag  + Q_{susy}^\dag {Q_{susy}}} \right)\end{equation}
We found that the L-BFGS-B optimizer again yielded the best results for the ground state energy. Using the VQE and 5 qubits ( 4 for the boson and 1 for the fermion) we obtain for the no noise simulation the convergence graph in figure 4 and the results in table 3 and for the noisy simulation the convergence graph in figure 5 and the results in table 4. Again we found that the L-BFGS-B optimizer again yielded the best results for the ground state energy. The results for the no noise simulation were again very accurate but were slightly less accurate than the one boson formulation. The noisy simulation of the one boson one fermion formulation was the least accurate of the quantum computations of superconformal quantum mechanics.
\begin{figure}
\centering
  \includegraphics[width = .45 \linewidth]{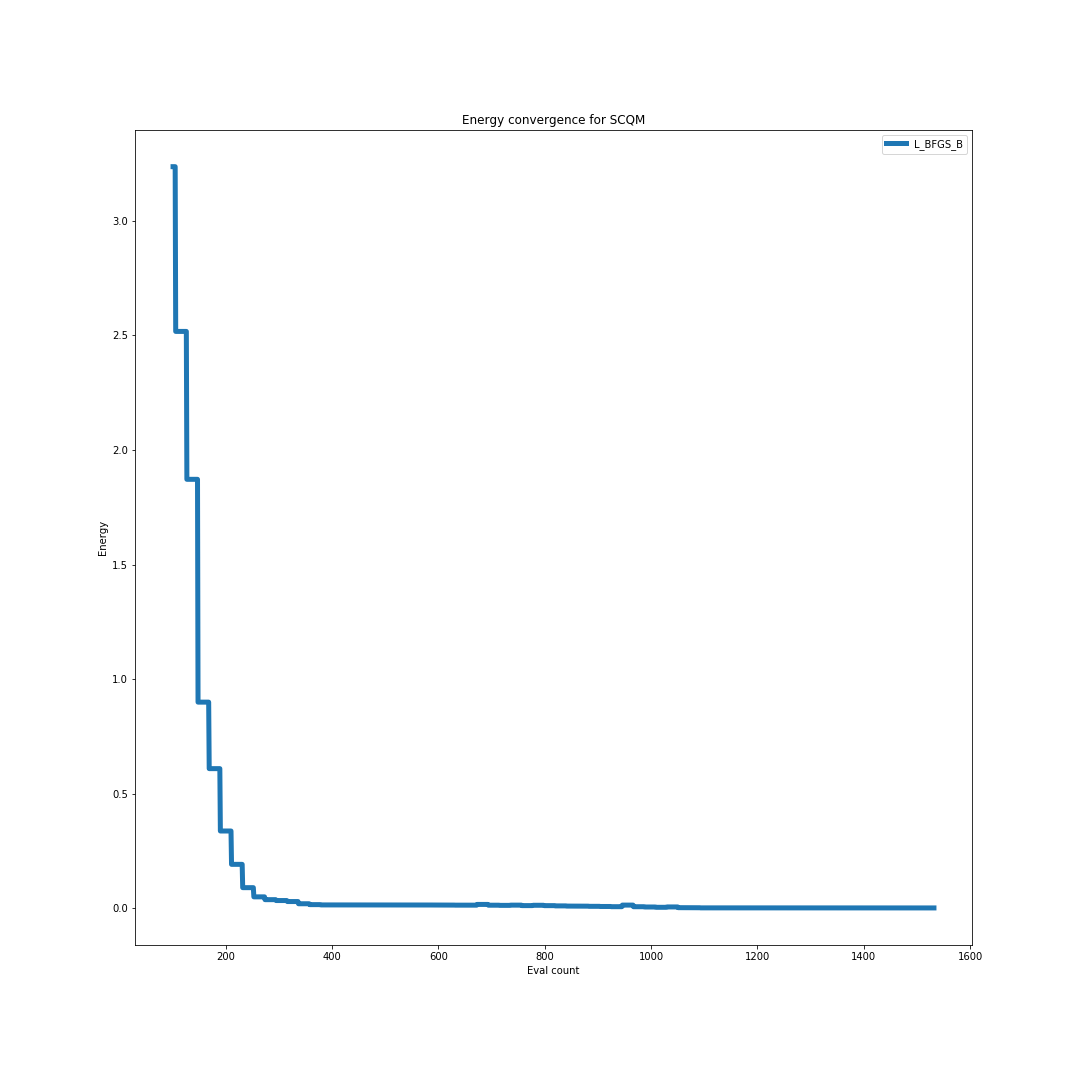}
  \caption{Convergence graph for the VQE computation of the ground state energy of SCQM using the one boson one fermion formulation. The VQE result was obtained using the State-vector simulator with no noise and the Limited-memory BFGS Bound (L-BFGS-B)  optimizer.}
  \label{fig:Radion Potential}
\end{figure}
\begin{figure}
\centering
  \includegraphics[width = .45 \linewidth]{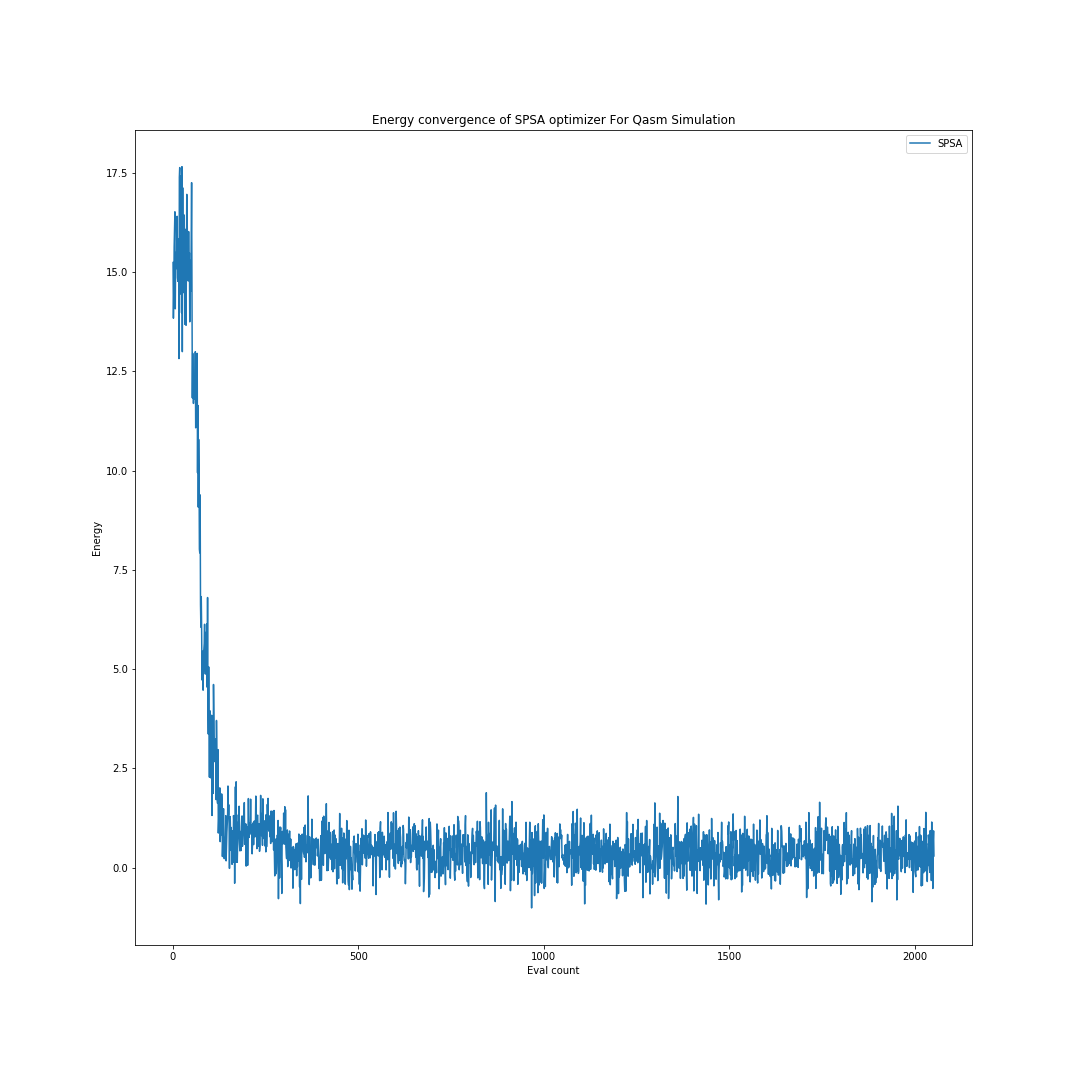}
  \caption{Convergence graph for the VQE computation of the ground state energy of SCQM using the one boson one fermion formulation. The VQE result was obtained using the QASM simulator including realistic noise and the Simultaneous Perturbation Stochastic Approximation (SPSA) optimizer.}
  \label{fig:Radion Potential}
\end{figure}
\begin{table}[h]
\centering
\begin{tabular}{|l|l|}
\hline
SCQM one boson one fermion formulation       & Ground State Energy \\ \hline
Exact   &    0.0        \\ \hline
Exact Discrete   &   0.00190358       \\ \hline
VQE-State-vector (L-BFGS-B optimizer)   &   0.00190402        \\ \hline
\end{tabular}
\caption{\label{tab:table-name} Exact, Exact discrete, and VQE computation of the ground state energy of SCQM using the one boson one fermion formulation. The VQE result was obtained using the State-vector simulator with no noise and the Limited-memory BFGS Bound (L-BFGS-B)  optimizer.}
\end{table}
\begin{table}[h]
\centering
\begin{tabular}{|l|l|}
\hline
SCQM one boson one fermion formulation       & Ground State Energy \\ \hline
Exact   &    0.0        \\ \hline
Exact Discrete   &   0.00190358       \\ \hline
VQE-QASM-Simulator (SPSA optimizer)   &   0.29056190       \\ \hline
\end{tabular}
\caption{\label{tab:table-name} Exact, Exact discrete, and VQE computation of the ground state energy of SCQM using the one boson one fermion formulation. The VQE result was obtained using the QASM simulator including realistic noise and the Simultaneous Perturbation Stochastic Approximation (SPSA) optimizer.}
\end{table}

\section{Evolution of Hamiltonian and Kernel for SCQM on a quantum computer}

The Kernel or Greens function for a quantum system tells how a system evolves from one state to another as a function of time. Like the energy spectrum it is a fundamental quantity to compute for any quantum system. It is given for position states by \cite{Feynman}:
\begin{equation}K\left( {x,x';T} \right) = \left\langle {x'\left| {x,T} \right\rangle } \right. = \left\langle {x'} \right|{e^{ - iHT}}\left| x \right\rangle  = \int_{x(0) = x}^{x(T) = x'} {Dx{e^{i\int_0^T {L\left( {x,\dot x} \right)dt} }}} \end{equation}
Where $H$ is the Hamiltonian and $L$ is the Lagrangian. For Superconformal quantum mechanics (ignoring constant terms) this is given by:
\[H = \frac{1}{2 m}{p^2} + \frac{{{m ^2\omega ^2}}}{2 m}{x^2} + \frac{{{g^2-g}}}{{{2 m x^2}}}\]
\begin{equation}L\left( {x,\dot x} \right) = \frac{m}{2}{{\dot x}^2} - \frac{{{m ^2 \omega ^2}}}{2 m}{x^2} - \frac{{{g^2-g}}}{{{2 m x^2}}}\end{equation}
 Like the energy spectrum for Superconformal quantum mechanics the kernel or Greens function for superconformal quantum mechanics can be computed exactly. The result is given by \cite{Schulman:1981vu}:
\begin{equation}K\left( {x,x';T} \right) = \frac{{m \omega \sqrt {xx'} }}{{i\hbar \sin \omega T}}\exp \left\{ {\frac{{i m\omega }}{{2\hbar }}\left( {{x^2} + x{'^2}} \right)\cot \omega T} \right\}{I_\gamma }\left( {\frac{{m \omega xx'}}{{i\hbar \sin \omega T}}} \right)\end{equation}
with
\begin{equation}\gamma  = \frac{1}{2}\sqrt {1 + \frac{{8 m(g^2-g)}}{{{2 m\hbar ^2}}}} \end{equation}
and where $I$ is a modified Bessel function. The Evolution of Hamitonian (EOH) quantum algorithm is based on the quantum walks quantum algorithm.  The quantities computed are the transition probability from one position to the same or another position as a function of time which defines the Kernel or propagator \cite{Feng:2021whf}.  We use the Qiskit EOH quantum algorithm to calculate the evolution of Hamiltonian. To construct the Hamiltonian we use the finite difference basis discussed in section 2 where the position and momentum matrices are represented by sparse matrices.

The Hamiltonian evolution is approximated using the Trotter-Suzuki decomposition. The simplest Trotter-Suzuki decomposition is given by
\begin{equation}e^{x\left(A+B\right)}=e^{xA}e^{xB}+O\left(x^2\right)\end{equation}
where x is a parameter and A and B are arbitrary operators with some commutation relation \( [A, B] \ne 0\).
To deal with higher order case, we need more parameters to do the corrections, with the generalized form of Trotter formula as 
\begin{equation}e^{x\left(A+B\right)}=e^{p_1xA}e^{p_2xB}e^{p_3xA}e^{p_4xB}\cdots e^{p_MxB}+O\left(x^{m+1}\right)\end{equation}
The set of parameters \(\lbrace p_1,p_2,\cdots,p_M\rbrace\) corresponding to order of \(x^{m+1}\).
%
%
%
%
To calculate time evolution over the time slice as \(t/n\) of a Hamiltonian system with the kinetic energy \(K(p)\), we could use the Trotter decomposition approximates the time evolution with the operator
\begin{equation}e^{itH}\simeq{\left(e^{\frac{it}{n}K}e^{\frac{it}{n}\nu}\right)}^n\end{equation}

\subsection*{Implementations on IBM quantum computers}

We use Qiskit to build the circuits for the EOH algoritm and simulate the time evolution. First the Hamiltonian matrix operator is converted to an expansion of Pauli terms that can read into the quantum computer.  
A larger number of time slices in each calculation would give a more accurate result, especially for a longer time interval. We used the Trotter-Suzuki expansion for all the calculations, with the expansion order equal to 3. We used the state vector simulator to run the EOH circuit. Each run of the circuit  returns a complex vector as a result, which can be used to define the probability distribution. To get a series results, we used a for-loop in the code to return the results corresponding to different time intervals. In this way, we see how system is evolving step by step in a long time interval. 

The results from the EOH quantum algorithm for Superconformal Quantum Mechanics are shown in figure 6 and for a time evolution from one position back to the same position or other positions. We were not able to take the evolution out far enough to exhibit the periodicity in $T$ that is present in the exact solution (6.3). Nevertheless the time evolution compares well with the exact time evolution for smaller time intervals that we plot in figure 7. The use of large matrices and more qubits together with smaller time slices should improve the representation of time evolution using the quantum computer.
%

\begin{figure}[!htb]
\centering
\minipage{0.4\textwidth}
  \includegraphics[width=\linewidth]{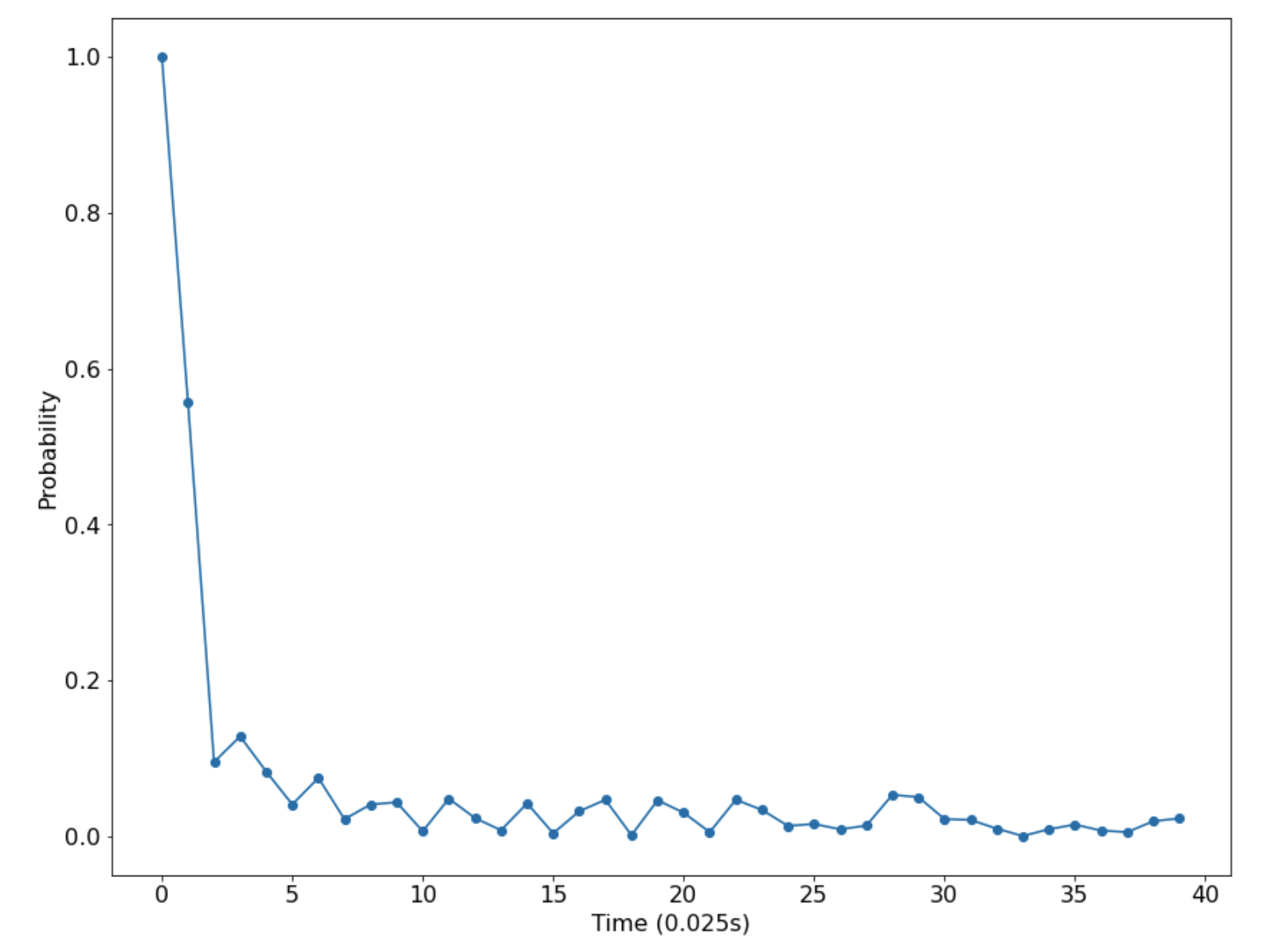}
\endminipage\hfill
\minipage{0.4\textwidth}
  \includegraphics[width=\linewidth]{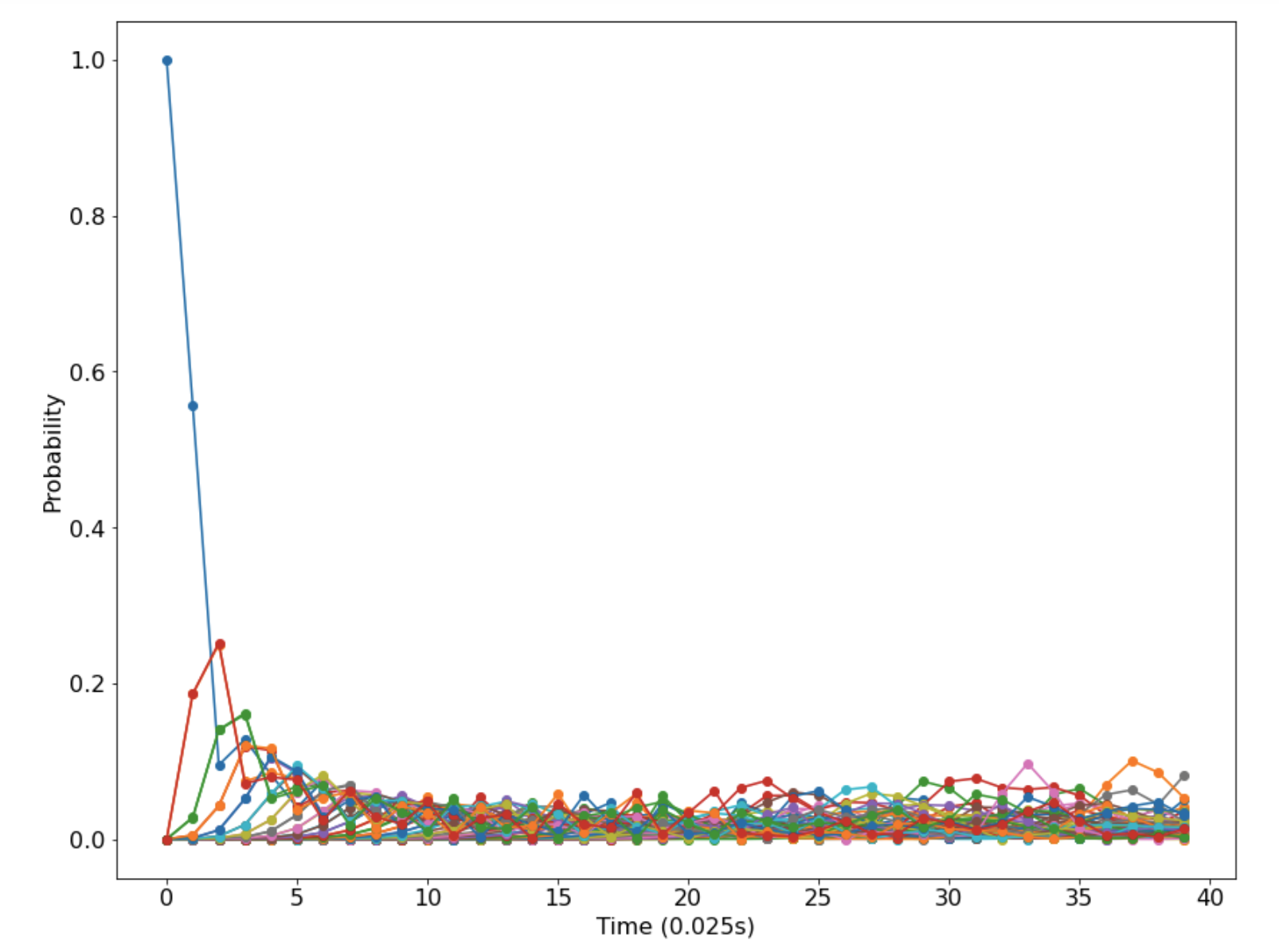}
\endminipage\hfill
\caption{(Left) Result of Evolution of Hamiltonian quantum algorithm for Supersymmetric Conformal Quantum Mechanics for $g=\sqrt{2}$ from one position back to the same position. (Right) Result of Evolution of Hamiltonian quantum algorithm for Supersymmetric Conformal Quantum Mechanics for $g=\sqrt{2}$ from one position back to the same position or another position.}
\end{figure}
  \label{fig:Radion Potential}
\begin{figure}
\centering
  \includegraphics[width = .4 \linewidth]{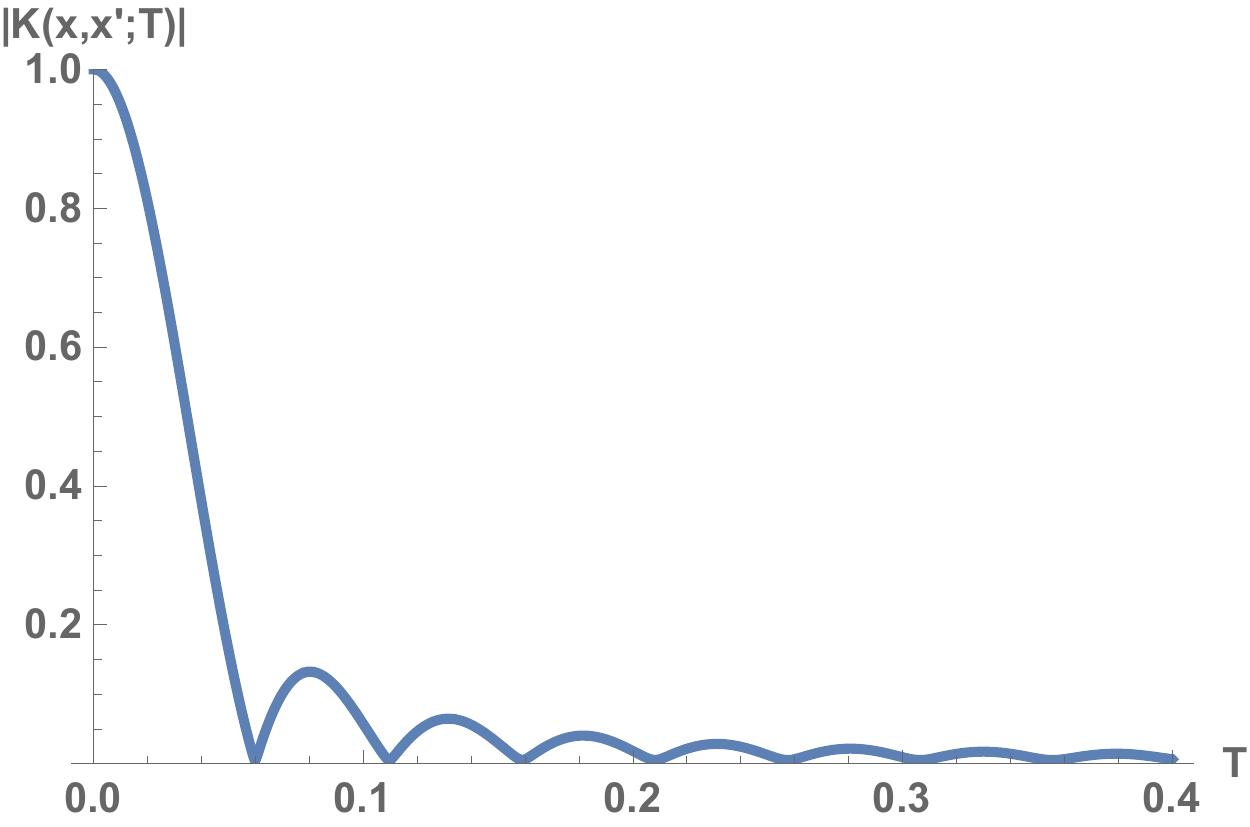}
  \caption{Exact solution for the time evolution of the Hamiltonian of Supersymmetric Conformal Quantum Mechanics potential for $g=\sqrt{2}$ form one position back to the same position.}
  \label{fig:Radion Potential}
\end{figure}

\section{Bosonic Calogero-Moser-Sutherland Model  on a Quantum Computer}

The bosonic Calogero-Moser-Sutherland model is a generalization of conformal quantum mechanics to a system with $N$ bosons. It is defined by the Hamiltonian \cite{Freedman:1990gd}:
\begin{equation}{H_B} = \frac{1}{2}\sum\limits_{i = 1}^N {p_i^2}  + {V_B}\left( {{x_1},{x_2}, \ldots ,{x_N}} \right)\end{equation}
with
\begin{equation}{V_B}\left( {{x_1},{x_2}, \ldots ,{x_N}} \right) = \frac{{{\omega ^2}}}{{2N}}\sum\limits_{i < j} {{{\left( {{x_i} - {x_j}} \right)}^2}}  + {g^2}\sum\limits_{i < j} {{{\left( {{x_i} - {x_j}} \right)}^{ - 2}}} \end{equation}
The ground state energy can be determined exactly and is given by:
\begin{equation}{E_0} = \frac{1}{2}\left( {N - 1} \right)\left( {1 + N\alpha } \right)\omega \end{equation}
with
\begin{equation}\alpha  = \frac{1}{2} + \sqrt {\frac{1}{4} + {g^2}} \end{equation}
For the special case $N=3$ the Hamiltonian becomes:
\begin{equation}H = \frac{1}{2}\left( {p_1^2 + p_2^2 + p_3^2} \right) + \frac{\omega^2}{6}\left( {{{({x_{12}})}^2} + {{({x_{23}})}^2} + {{({x_{31}})}^2}} \right) + {g^2}\left( {{{({x_{12}})}^{ - 2}} + {{({x_{23}})}^{ - 2}} + {{({x_{31}})}^{ - 2}}} \right)\end{equation}
where
${x_{ij}} = {x_i} - {x_j}$. Here we take $N=3$, $\omega = 1$ and $g= \sqrt 2$ so that $\alpha = 2$ and the exact value $E_0 = 7$.
To perform the quantum computation of the bosonic CMS model we construct the position and momentum operator matrices. Here we choose the position basis and represent each boson by a $4\times 4 $ matrix. Defining the position
$$ X_p = \left(
 \begin{array}{cccc}
 -\frac{3 \sqrt{\frac{\pi }{2}}}{2} & 0 & 0 & 0 \\
 0 & -\frac{\sqrt{\frac{\pi }{2}}}{2} & 0 & 0 \\
 0 & 0 & \frac{\sqrt{\frac{\pi }{2}}}{2} & 0 \\
 0 & 0 & 0 & \frac{3 \sqrt{\frac{\pi }{2}}}{2} \\
\end{array}
\right)$$
\begin{equation}
\end{equation}
and the momentum using the Sylvester matrix $F$ discussed in section 2 as
\begin{equation}{P_p} = {F^\dag }{X_p}F\end{equation}
We can construct the position and momentum matrices using the Kronecker product through:
\[{x_1} = {X_p} \otimes {I_4} \otimes {I_4}\]
\[{x_2} = {I_4} \otimes {X_p} \otimes {I_4}\]
\[{x_3} = {I_4} \otimes {I_4} \otimes {X_p}\]
\[{p_1} = {P_p} \otimes {I_4} \otimes {I_4}\]
\[{p_2} = {I_4} \otimes {P_p} \otimes {I_4}\]
\begin{equation}{p_3} = {I_4} \otimes {I_4} \otimes {P_p}\end{equation}
The total Hilbert space for the three bosons is then $4 \times 4 \times 4 = 64$ dimensional. We used the VQE algorithm to compute the ground state energy on the quantum computer using six qubits. For the bosonic CMS model we found that the L-BFGS-B optimizer yeilds the best results. The convergence graph for this optimizer in shown in figure 8. In table 5 we record the results of the quantum computation. The results were not as accurate as the single boson  SCQM or one boson and one fermion SCQM computations. This may be because the variational wave functions in the three boson case do not overlap with the true ground state as strongly for the bosonic CMS as in the SCQM case. It would be interesting to examine the true ground state in more detail and explore different variational ansatz for the trial wave functions to see if one can improve the accuracy of the result for the bosonic CMS model.
\begin{table}[h]
\centering
\begin{tabular}{|l|l|}
\hline
Bosonic CMS Model       & Ground State Energy \\ \hline
Exact   &    7        \\ \hline
Exact Discrete   &    6.59198266     \\ \hline
VQE (L-BFGS-B)   &   7.12370445       \\ \hline
\end{tabular}
\caption{\label{tab:table-name} Exact, Exact discrete, and VQE computation of the ground state energy of bosonic CMS model for $g = \sqrt 2$ using the position basis.  The VQE result was obtained using the State-vector simulator with no noise and the Limited-memory BFGS Bound (L-BFGS-B)  optimizer.}
\end{table}
\begin{figure}
\centering
  \includegraphics[width = .4 \linewidth]{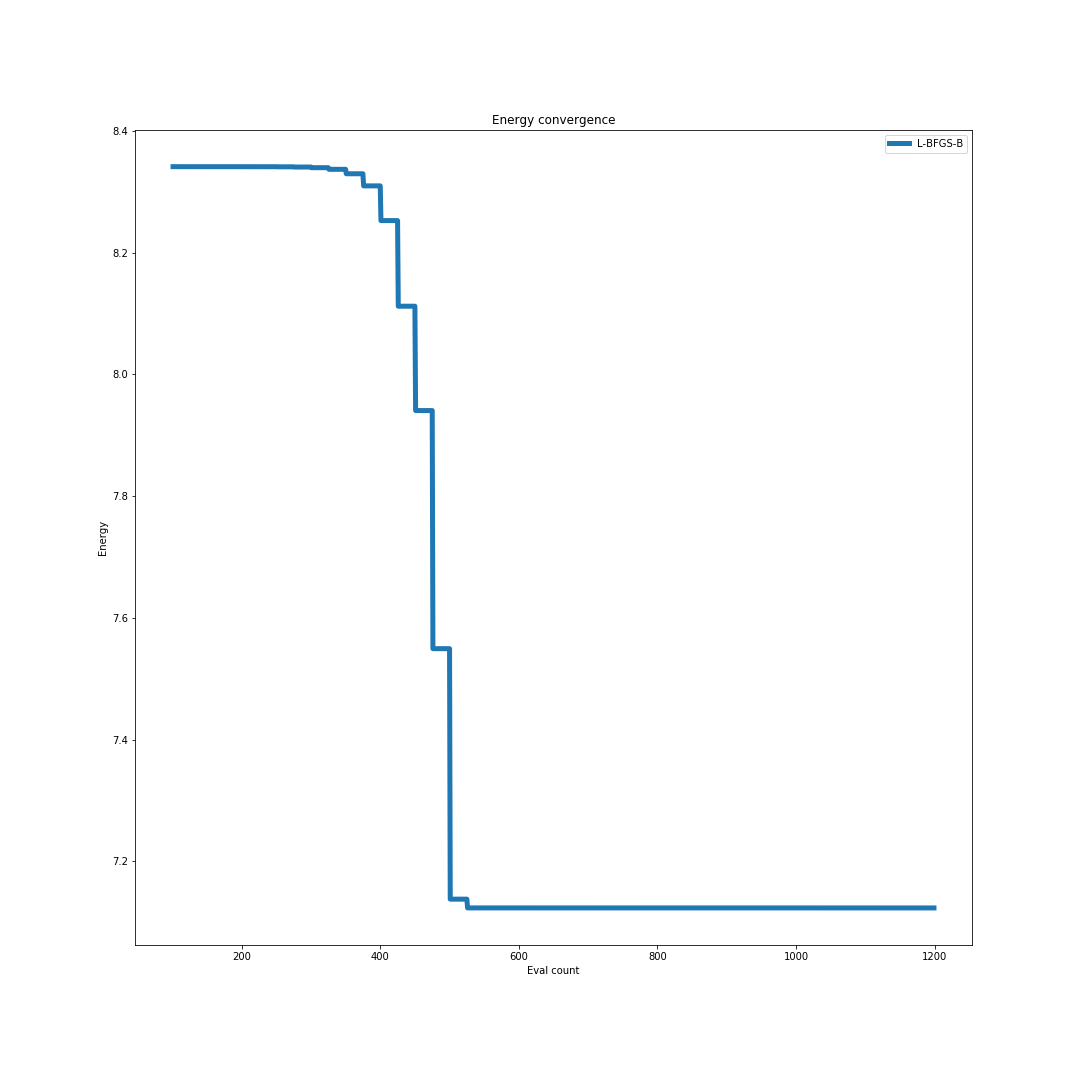}
  \caption{Convergence graph for the VQE computation of the ground state energy of  bosonic CMS model for $g = \sqrt 2$ using the position basis. The VQE result was obtained using the State-vector simulator with no noise and the Limited-memory BFGS Bound (L-BFGS-B)  optimizer.}
  \label{fig:Radion Potential}
\end{figure}
\newpage

\section{Quantum computing for Supersymmetric Calogero-Moser model}

The Supersymmetric Calogero-Moser-Sutherland model is a generalization of SCQM with $N$ bosons and $N$ fermions. It is defined by the Hamiltonian \cite{Freedman:1990gd}\cite{Desrosiers:2003me}:
\begin{equation}H = \frac{1}{2}\sum\limits_{i = 1}^N {\left[ {p_i^2 + {\omega ^2}x_i^2} \right]}  + \sum\limits_{1 \le i < j \le N} {\frac{{g(g - 1 + {\theta _{ij}}\theta _{ij}^\dag )}}{{x_{ij}^2}}}  - \frac{N}{2}\left[ {g(N - 1) - \omega } \right]\end{equation}
where
\[{x_{ij}} = {x_i} - {x_j}\]
\begin{equation}{\theta _{ij}} = {\theta _i} - {\theta _j}\end{equation}
Defining the supercharge $Q$ so that:
\begin{equation}H = \frac{1}{2}\left\{ {Q,{Q^\dag }} \right\}\end{equation}
where
\begin{equation}Q = \sum\limits_{j = 1}^N {{\theta _j}} \left( {i{p_j} + \omega {x_j} - g\sum\limits_{1 \le k \le N,k \ne j} {\frac{1}{{{x_{jk}}}}} } \right)\end{equation}
and 
\begin{equation}Q = \sum\limits_{j = 1}^N {{\theta _j}} \left( { - i{p_j} + \omega {x_j} - g\sum\limits_{1 \le k \le N,k \ne j} {\frac{1}{{{x_{jk}}}}} } \right)\end{equation}
The ground state satisfies $Q \psi_0 = 0$ and is given by:
\begin{equation}
{\psi _0}(x) = \left( {{{\prod\limits_{1 \le j < k \le N} {\left( {{x_{jk}}} \right)} }^g}} \right)\prod\limits_{i = 1}^N {{e^{ - \omega \frac{{x_i^2}}{2}}}} \end{equation}
To perform the quantum computation we again choose the position basis where each position and momenetum matrix is represented using 2 qubits so as $4 \times 4$ matrices. In addition we add three fermion matrices which use 1 qubit each so are represented by $2 \times 2 $ matrices. The entire Hilbert space is $4 \times 4 \times 4 \times 2 \times 2 \times 2 = 512 $ dimensional so uses 9 qubits total. This is the largest quantum computation studied in this paper. The position and fermion operators are represented as:
\[{x_1} = {X_p} \otimes {I_4} \otimes {I_4} \otimes {I_8}\]
\[{x_2} = {I_4} \otimes {X_p} \otimes {I_4}\otimes {I_8}\]
\[{x_3} = {I_4} \otimes {I_4} \otimes {X_p}\otimes {I_8}\]
\[{\theta _1} = {I_{64}} \otimes \left( {\begin{array}{*{20}{c}}
0&1\\
0&0
\end{array}} \right) \otimes {I_2} \otimes {I_2}\]
\[{\theta _2} = {I_{64}} \otimes {\sigma _3} \otimes \left( {\begin{array}{*{20}{c}}
0&1\\
0&0
\end{array}} \right) \otimes {I_2}\]
\begin{equation}{\theta _3} = {I_{64}} \otimes {\sigma _3} \otimes {\sigma _3} \otimes \left( {\begin{array}{*{20}{c}}
0&1\\
0&0
\end{array}} \right)\end{equation}
From these one can construct the Hamiltonian from (8.1). Using this Hamiltonian for the VQE we find that the Constrained Optimization By Linear Approximation (COBYLA) optimizer yielded the best results which we list table 6. The reason the exact discrete vacuum energy is not zero is due to to the approximation of infinite dimensional position and momentum matrices by finite matrices. This discretization error can be reduced by going to higher number of qubits.
%
%
\begin{table}[h]
\centering
\begin{tabular}{|l|l|}
\hline
Susy CMS Model       & Ground State Energy \\ \hline
Exact   &    0.0        \\ \hline
Exact Discrete   &    -0.04716555       \\ \hline
VQE (COBYLA)   &   0.75686596        \\ \hline

\end{tabular}
\caption{\label{tab:table-name} Exact, Exact discrete, and VQE computation of the ground state energy of supersymmetric CMS model for $g = \sqrt 2$ using the position basis.  The VQE result was obtained using the State-vector simulator with no noise and the Constrained Optimization By Linear Approximation (COBYLA)  optimizer.}
\end{table}
\newpage

\section{Conclusions}

In this paper we have investigated the simplest superconformal field theory in 0+1 dimension on a quantum computer which is superconformal quantum mechanics (SCQM). We studied the ground state of the mass deformed SCQM on a quantum computer using the Variational Quantum Eigensolver (VQE) quantum algorithm using a one boson and a one boson - one fermion Hilbert space and compared the results. We studied the Feynman path integral for SCQM using the EOH algorithm on the quantum computer using the Trotter-Suzuki approximation and compared with the exact result.  Finally we considered an N boson and N fermion version of SCQM given be the Supersymmetric Calogero-Moser model. We compared the ground state of theory obtained using the VQE computation with the exact solution known for the theory. The accuracy of the one boson formulation of SCQM was the best, followed by the one boson  - one fermion formulation, bosonic Calogero-Moser-Sutherland model and the Supersymmetric Calogero-Moser Sutherland model was the least accurate. Improved variational ansatz should improve the accuracy of the VQE computations for the multiparticle simulations as it does for quantum chemistry simulations of molecules. For the time evolution finer divisions of the time interval and higher order expansions used in the Trotter-Suzuki approximation should improve the accuracy, but at the cost of longer and more complex quantum circuits used in the computations.

We emphasize that superconformal field theories, including SCQM which can be considered a superconformal field theory in 0+1 dimensions, are thought to be dual to a Universe with negative cosmological constant or Anti-de Sitter space (AdS)  due to the AdS/CFT correspondence \cite{Maldacena:1997re}. Thus if we are simulating a superconformal field theory we can, also in the same computation, simulating a Universe \cite{Tegmark:2007ud}
\cite{Beane:2012rz}, albeit one with negative cosmological constant. Furthermore the reduced computational complexity of simulating the superconformal field theory instead of the full quantum gravity may cause one to rethink the computational complexity of gravitational simulations \cite{Lloyd} \cite{Lloyd:2005js} and how one can simulate the Universe on a quantum computer \cite{Liu:2020eoa}\cite{Liu:2020wtr}\cite{Gharibyan:2020bab}.

\section*{Acknowledgements}

We thank Junyu Liu, Hans Guenter Dosch and Vladimir Akulov for sending additional references and useful comments on the paper.

\end{document}